\documentclass[review]{elsarticle}

\usepackage{graphicx}

\usepackage{amssymb}
\usepackage{amsmath}
\usepackage{natbib}

\journal{New Astronomy}

\begin{document}

\begin{frontmatter}



\title{The role of dark matter in the galaxy mass-size relationship}


\author[rvt]{D. Bindoni}
\ead{daniele.bindoni@unipd.it}
\author[rvt]{L. Secco}
\ead{luigi.secco@unipd.it}
\author[els]{E. Contini}
\author[rvt]{R. Caimmi}
\ead{roberto.caimmi@unipd.it}
\address[rvt]{Department of Astronomy, University of Padova, Padova, Italy}
\address[els]{Astronomical Observatory of Trieste, Trieste, Italy}

\begin{abstract}
{The observed relationship between stellar mass and effective radius for early type galaxies, pointed out by many authors, is interpreted in the context of Clausius' virial maximum theory. 
In this view, it is strongly underlined that the key of the above mentioned correlation is owing to the presence of a deep link between cosmology and the existence of the galaxy Fundamental Plane. Then the ultimate meaning is: understanding visible mass - size correlation and/or Fundamental Plane means understanding how galaxies form.   
The mass - size relationship involves baryon (mainly stellar) mass and its typical dimension related to the light, but it gets memory of the cosmological mass variance at the equivalence epoch.  The reason is that the baryonic component
virializes by sharing virial energy in about equal amount between baryons and dark matter, this sharing depending, in turn, on the steepness of the dark matter distribution. 
The general strategy consists in using the two-component tensor virial theorem for determining the virialized baryonic configurations.
A King and a Zhao density profile are assumed for the inner baryonic and the outer dark matter component, respectively, at the end of the relaxation phase. All the considerations are restricted to spherical symmetry for simplicity.  The effect of changing the dark-to-baryon mass ratio, $m$, is investigated inside a $\Lambda$CDM scenario. 
A theoretical mass - size relation is expressed for the baryonic component, which fits fairly well to the data from a recently studied galaxy sample. Finally, the play of intrinsic dispersion on the mass ratio, $m$, is discussed in the light of the {\it cusp/core problem} and some consequences are speculated about the existence of a limit $m_l$ expected by the theory.}
\end{abstract}


\end{frontmatter}   
   
\section{Introduction}

The physical grounds on the basis of Clausius' virial maximum theory (CVMT) lies in that: to explicitely consider the additional degree of freedom the problem has, owing to the existence of one further component really present in a 
galaxy structure: the dark matter halo. This is not to be included into the system in a way to handle it as a single one and then applying the one-component tensor virial theorem. On the contrary, the system is splitted into two subsystems: one of baryons, the other one of non-baryonic dark matter (DM) to which the two-component tensor virial theorem is applied. Then the double system gains some new ways to share out the amount of potential energy which has to appear in the two virial equilibrium equations. Briefly summarizing, the general strategy consists to use the two-component tensor virial theorem (e.g., Brosche et al., 1983; Caimmi \& Secco, 1992) to describe the virial configuration of the baryonic component embedded in a DM halo at the end of relaxation phase (see, Bindoni \& Secco, 2008).
It reads:
\begin{eqnarray}
\label{vireq} 
2(T_{u})_{ij}=(V_{u})_{ij};\ \  (u=B,D; i,j= x,y,z)
\end{eqnarray}
According to the scalar virial for one component, the potential energy tensor, which has to enter into the tensor virial equations, is the Clausius' virial (CV) tensor, $(V_{u})_{ij}$, built-up of the \textit{self potential-energy tensor}, $(\Omega_u)_{ij}$, and the \textit{tidal potential-energy tensor}, $(V_{uv})_{ij}$ ($u=B,D$; $v=D,B$).
Then, according to the scalar virial theorem, the trace of CV tensor, related to the inner bright, B, component, has to be read: 
\begin{eqnarray}
\label{Clau}
V_B= \Omega_B + V_{BD}\\
\label{omB}
\Omega_B= \int\rho_B\sum_{r={1}}^{3}~ x_r\frac{\partial\Phi_B}{\partial x_r}{\rm\,d}\vec{x_B}~= \int\rho_B
(\vec{r_B}\cdot\vec{f_B}){\rm\,d}\vec{x_B}\\
\label{VBA}
(V_{BD})= \int\rho_B\sum_{r={1}}^{3} x_r\frac{\partial\Phi_D}{\partial x_r}{\rm\,d}\vec{x_B}~=\int\rho_B(\vec{r_B}\cdot\vec{f_D}){\rm\,d}\vec{x_B};
\end{eqnarray}
where $\rho_{B}$ is the $B$ component density and $\vec{f}_{B}, \vec{f}_{D}$ are the force per unit mass due to the  self and DM gravity, respectively, at the point $\vec{r}_B$ and  $\Phi_B$, $\Phi_D$ are the related potentials.
Conversely, the total potential energy tensor of the $B$ component is:
$(\Omega_B)_{ij} + (W_{BD})_{ij}$,
where the interaction energy tensor is: $(W_{BD})_{ij}=-\frac12\int\rho_{B}(\Phi_{D})_{ij}{\rm\,d}\vec{x}_{B}$; and the potential tensor (e.g., Chandrasekhar, 1969) due to the DM is:
$(\Phi_{D})_{ij}=G \int\rho_{D}(\vec{x^{\prime}})\frac{(x_{i}-x^{\prime}_{i})( x_{j}-x^{\prime}_{j})}{\mid{\vec{x}-\vec{x^{\prime}}\mid}^3}{\rm\,d}\vec{x}_{D}$.
 
To be noted that in general: $(V_{BD})_{ij}\neq(W_{BD})_{ij}$, the difference gives the residual energy tensor (Caimmi \& Secco, 1992).
 
In spite of the total potential energy: $(E_{pot})_B= \Omega_B+ W_{BD}$, which has an increasing monotonic trend by increasing the ratio between the volume taken by baryons over that of DM halo (Secco, 2005, hereafter quoted as LS5, Fig. 2), the trace of CV tensor, $V_B$, shows, under some constraints, a maximum (CVM) for a special virial configuration. That means, in turn,  a minimum of the random kinetic energy to obtain equilibrium for the baryonic component when it is completely embedded into a DM halo assumed with a fixed density profile and dimension.
The possibility to get a maximum is subjected to two restrictions, namely (1) the DM density distribution is decreasing not too fast inside the bulk of baryonic matter, and (2) the total DM mass maintains above a threshold.
Indeed we are dealing with an amount of CV energy to be divided between two components: that in baryons and that in DM fraction inside the baryon container (due to Newton's first theorem).
If a power-law DM density, $\rho_D\sim 1/r^d$, increases too fast toward the center, the weight of tidal energy tends to overcome that of the self gravitational energy due to baryons, until a divergence in the energy ratio occurs as soon as $d\rightarrow2$ (Marmo \& Secco, 2003). To explain some of the main Fundamental Plane (FP) features in the light of the CVMT theory, a DM distribution characterized by $d\simeq 0.5$ is needed (e.g., Bindoni, 2008).

Moreover, decreasing the DM total amount the sharing of virial energy needs that the DM fraction inside baryon container has to become all the available DM amount. In other words the radius corresponding to the CVM moves outwards so that the two volumes coincide. Under this thresold CVM doesn't exist.

In sect. 2 the existence of a special tidal dimension for the baryonic component due to the gravitational interaction of DM is underlined and a model to link it to the light dimension $r_e$ is described. At sect. 3 the $\Lambda$CDM cosmological scenario is introduced and the typical CVMT scaling laws of $r_e, I_e, \sigma_o$ with the only two free parameters $M_B, m$ are considered by means of the only two exponents: the slope of DM density profile and the cosmological local slope of mass variance. At sect. 4 and 5 theoretical fits vs. observations are taken into account with some linearization procedure and some useful relationships. In sect. 6 the theoretical slopes of mass-size relationship are successfully compared with those Tortora et al. (2009, hereafter quoted as Tal09) obtained by observations. The role of the intrinsic dispersion due to $m$ as function of the slope of DM density profile are discussed and compared with CVMT's expectations for some of the main scaling laws, in sects. 7,8,9. The discussion and conclusions follow at Section 10 and 11, respectively.

\section{Sharing virial energy between interacting components}

It is not surprising that a relationship between the visible mass, $M_B$\footnote{Mostly in stars, assumed to form very early. So the visible mass $M_B\simeq M_*$.} and its spatial dimension does exist for early type galaxies. Indeed, according to the CVM theory's expectation a special scale lenght is induced, when the maximum exists (see, Secco \& Bindoni, 2009, hereafter quoted as SB9), from the dark matter halo on the baryonic gravitational field. In this way the gravity, which for its nature has an infinity range, acquires an intrinsic macroscopic scale. Among the infinite possible dimensions corresponding to virial equilibrium of a $B$ component embedded inside a dark one, $D$, only one is able to maximize the Clausius' virial energy by sharing it in about equal amount between baryons and dark matter particles. The consequence is that visible virialized matter does not have a whatever dimension but a special one, that is the {\it tidal radius} $a_t$,  strictly related to that of DM virial radius, $a_D$.

The assumed DM density profile is a special case of Zhao (1996) family (for more insight see, Caimmi et al., 2005):
\begin{equation}
\label{dmdp}
\rho_D=\frac{2\rho_{oD}}{1+(\frac{r}{r_{oD}})^d}
\end{equation}
where $r_{oD}$ is a scale radius and $\rho_{oD}=\rho(r_{oD})$.
More specifically, the profile of Eq.(\ref{dmdp}) has a central core and tends to a power-law density profile with exponent $d$ for sufficiently large radial distances. We will come back to the relevance of the d-value in the next sub-section.

The related DM virial radius - tidal B-radius relation is found to be (SB9):
\begin{equation}
\label{at}
a_{t}=\Big( \frac{\nu_{\Omega B}}{\nu'_V}
\frac{1}{2-d}\frac{M_B}{M_D}\Big)^{\frac{1}{3-d}} a_D
\end{equation}
where $a_t$, $a_D$, are the tidal and the virial radius, respectively, $M$ is the total mass, $\nu_{\Omega B}$ the
{\it self-mass distribution coefficient}, $\nu'_V$ the {\it reduced interaction coefficient}, and the indices, B, D, denote the star and DM subsystem, respectively.
The reduced interaction coefficient is defined as: 
\begin{equation}
\label{nu}
\nu'_V(x)=\frac{\nu_V(x)}{m x^{3-d}};~~x=\frac{a_B}{a_D};~~m=\frac{M_D}{M_B};
\end{equation}
where $\nu_V$ is the {\it interaction coefficient}, $x$ the B to D virial radius ratio, and $m$ the D to B total
mass ratio. It can be seen that $\nu'_V(x)\approx$ const to a first extent (SB9).
For deeper insight to Eq. (\ref{at}), further considerations are needed. A dimensionless trace of CV tensor can be obtained dividing both sides of Eq.(\ref{Clau}) by a normalization energy, $GM_B^2F/a_D$, where $G$ is the gravitational
constant and $F$ a form factor ($F$=2 in the case under discussion of spherical-symmetric configurations).
The result is:
\begin{equation}
\label{norvi}
\tilde{V}_B=-\frac{\nu_{\Omega B}}{ x}-\frac{\nu_V(x)}{x}.
\end{equation}
The condition: $\frac{d\tilde{V_B}}{dx }=0$, yields the value $a_t$ expressed by Eq.\,(\ref{at}).
It is noteworthy that existence of $a_t$ which means from one side the existence of the maximum in CV energy and then its sharing between the two components, implies from the other side the existence of a minimum value for $m$. Indeed, if the baryonic component has to be embedded inside the DM halo (according to the tensor virial theorem extended to two componens) the location limit of CVM is the DM component border. Studing how changes the ratio $x_t=a_t/a_D$ as $m$ changes it is easy to prove that $x_t$ grows at decreasing $m$ so that $a_t$ reaches $a_D$ for the lowest value:
\begin{equation}
\label {ml}
m_l=\frac{\nu_{\Omega B}}{\nu'_V}\frac{1}{(2-d)}
\end{equation}
corresponding to $x_t=1$ in Eq.\,(\ref{at}).
That in turn leads to the expectation of zones of exclusion in the scaling laws based on the CVM presence. 
In fact, values $m<m_l$ would imply a DM halo embedded within the star subsystem, contrary to both current cosmological scenarios and observations such as flat rotation curves in galactic disks which extend well outside the visible region.

It is to underline the strong analogy which exists between the {\it tidal radius} $a_t$ and that von Hoerner (1958) found for a spherical star cluster embedded in the Galaxy tidal potential. At first sight the result might appear surprising because in the cluster case we are dealing with two forces, the self gravitational- and  tidal-force which are acting in opposite directions; here, on the contrary, the {\it tidal radius} is a consequence of two attractive forces toward the same central direction: the self gravity per unit mass $\vec{f}_{B}$ and the other one $\vec{f}_{D}$ due to DM  both applied to each point $\vec{r}_B$ (Eqs.\ref{omB},\ref{VBA}).
But the balance which defines $a_t$ is not between forces but between energies, given by forces times positions: {\it self} and {\it tidal}. They have different trends as $a_B$ decreases.

Tidal energy of $B$ due to subsystem $D$, goes as follows: according to first Newtons' theorem the fraction of $D$ mass which exerts dynamical effect on the $B$ component is only that inside the $B$ boundary, which for spherical-symmetric configurations is proportional to $M_D (\frac{a_B}{a_D})^{3-d}$, then the integral of force times the position 
decreases toward $0$ like $M_D (\frac{a_B}{a_D})^{2-d}$ at decreasing $a_B$.
The other limit that is the maximum value of such an energy, corresponds to the configuration in which the whole $D$ mass is inside the $B$ component i.e. the two boundaries coincide. Exactly the contrary occurs to the self energy: it grows as $a_B$ decreases and tends to zero at increasing $a_B$.

Then in a configuration where $a_B~>~a_t$, the tidal energy term in the Clausius' virial (Eq.\ref{Clau}) is dominant in respect to the self-energy one. 
When $a_B$ becomes less than $a_t$, it occurs the overwhelming of the self-energy over tidal-energy. All that turns to be in strict analogy not only with von Hoerner's radius but also with the Hill's radius (see, Binney \& Tremaine, Chapt.8, 2008). Indeed on the basis of both definitions it lies the restricted three body problem.  
But it is to be noted that our new definition of {\it tidal radius}\  (Eq.\ref{at}) refers not to two off centre objects but to two concentric ones. Then it looks like a generalization of those previously given.

\subsection{CVMT and DM cusp/core problem}

Whithout entering the complex, still open problem (see, e.g., Bindoni, 2008) we underline only the relevance it has inside the CVMT.
With regard to DM density profiles characterized by the presence of a central cusp, such as NFW (Navarro et al., 1996), the term ''cored'' has to be intended as related to milder $(\rho_{DM}\propto r^{-1/d},~d<1)$ slopes in the central region instead of erasing the cusp. A similar situation occurs in the case under consideration.
With regard to DM density profiles characterized by the presence of a central core, such as the one expressed by Eq.(\ref{dmdp}), a ''cuspy'' behaviour is shown by steeper $(d\ge1)$ slopes at sufficiently large $(r>>r_{oD})$ radial distances without erasing the core. What is relevant is indeed the slope of DM inside which the bulk of baryonic matter in stars is embedded because the gravitational interaction may lead for the two cases: $d < 1$, $d \ge 1$, to completely different results.
%
%
Coming back to the normalized virial energy (\ref{norvi}) and getting advantage by the approximation $\nu_V'\approx$ const, see Eq.\,(\ref{nu}), the ratio $\zeta_t$ between the self and the tidal energy at $x_t$, becomes (Marmo \& Secco, 2003):
\begin{equation}
\label{zita}
\zeta_t=\frac{\tilde{\Omega}_B}{\tilde{V}_{BD}} \simeq (2-d)
\end{equation}
which means the exact balance between self and tidal contibution is reached when the logarithmic slope of the DM density profile is $d=1$ (i.e., cuspy profile). Then $\tilde{V}_{BD}/\tilde{\Omega}_B$ increases towards $\infty$ as soon as $d \rightarrow 2$.
Moreover, it is easy to prove that in this cuspy case, $d$=1, the contribution of DM tidal component to virial energy becomes of the same amount of the baryonic one, whichever is the ratio $m$. In other words this case is equivalent to handle with a unique component in the virial equilibrium equation having an  equivalent  self-mass distribution coefficient,
$\bar{\nu}_{\Omega}=2\cdot{\nu}_{\Omega B}$, two times that of the single baryonic one. It means that a double system considered on the maximum of its inner component virial energy may degenerate into a single system only in this sense: not if DM mass reduces to $0$ (because the DM amount may be reduced only until to $m_l$) but as soon as the DM component is able to share exactly with the inner one the amount of its virial energy.

\subsection{The $r_{e}-a_{t}$ link}

The relationship $r_e-M_*$ can be obtained via Eq.(\ref{at}), which links the size ratio between the two components to their mass ratio, provided a connexion exists between $r_e$ and the {\it tidal radius} $a_t$, by a model, within a cosmological scenario where the DM halo dimension is related to the density primordial perturbation spectrum.

We assume that the ETGs are members of a perfect {\it homologous family} and then for each galaxy we handle with three different profiles for baryonic, light and DM distributions, each of them assumed to be the same for the whole family. The universality of our choice might be questionable in particular that of DM either for the still pending cusp/core problem or for the difficulty to reformulate the CV theory when baryons are embedded inside a halo density profile very different from that of a power-law. 

Here we refer to our non-linear spherical model considered in SB9 where it was assumed for the bright B-component the empirical surface light density law proposed by King (1962) as:
\begin{equation}
\label{king}
I(R)=k_L\left\{\frac{1}{[1+(R/R_c)^2]^{1/2}}-\frac{1}{[1+(R_t/R_c)^2]^{1/2}}\right\}^{2}
\end{equation}
where $R_t$ is the value of projected radius\footnote{Due to the choice of spherical model the projected radial quantities: $ R_t, R_c, R_e$, coincide with the corresponding three dimensional ones: $r_t, r_c, r_e$.} $R$ at which
$I$ reaches zero.
The law has the advantage to take into account the existence of a cut-off in the surface light distribution as one expects for globular clusters (GC) (the profile (\ref{king}) was born indeed for GCs and then it has been extended to ETGs). As shown in previous subsections, that is particular suitable because we expect for ETGs the presence of a cut-off which is $a_t$ and has an analogous role of von Hoerner's one.

The value $R_c$ corresponds to the core-radius and the $k_L$ value is linked to the central surface light density $I_o$ by:
\begin{equation}
\label{fo}
I_o=k_L\left\{1-\frac{1}{[1+(R_t/R_c)^2]^{1/2}}\right\}^{2}
\end{equation}
The corresponding baryonic matter projected density is:
\begin{equation}
\label{masP}
\Sigma (R)=k_{M}\left(\frac{1}{[1+(\frac{R}{R_c})^2]^{1/2}}
-\frac{1}{[1+(\frac{R_t}{R_c})^2]^{1/2}}\right)^2
\end{equation}
which is linked to the spatial baryonic matter density by the Abel integral equation (Binney \& Tremaine, 1987, Chapt.4):
\begin{equation}
\label{abel}
\rho(r)=-\frac{1}{\pi}\int_r^{r_t}\frac{{\rm\,d}\Sigma}{{\rm\,d}R}\frac{{\rm\,d}R}{\sqrt{R^2-r^2}}
\end{equation}
It should be noted that in our model $k_M \ne k_L$ being one of our main assumptions (see, SB9). 
Integration of $\Sigma (R)$  with respect of $2\pi R {\rm\,d}R$ gives the total projected mass within the projected distance $R$ from the center which becomes the luminosity function, $L(X)$, with the substitution of $\Sigma (R)\rightarrow I(R)$ and $k_M\rightarrow k_L$:
\begin{equation}
\label{limL}
L\left(X\right)=\pi r^{2}_{c}k_{L}F_L(X)
\end{equation}
where:
\begin{eqnarray}
\label{flx}
F_L(X)=\left[\ln\left(1+X\right)-4\frac{\left(1+X\right)^{1/2}-1}{\left(1+X_{t}\right)^{1/2}}+\frac{X}{1+X_{t}}\right]\\
X=\left(\frac{R}{r_{c}}\right)^{2}\ \ \ ,\ \ \ X_{t}=\left(\frac{R_{t}}{r_{c}}\right)^{2}
\end{eqnarray}
According to Eq.(\ref{limL}), as $X_t>>1$ the limit of $L(X_t)$ goes approximately to:
\begin{equation}
\label{LL}
L\left(X_t\right)\simeq\pi r^{2}_{c}k_{L}\ \ln\left(\frac{r_t^2}{20 r_c^2}\right)
\end{equation}
The square of the effective radius $r_e$ normalized to $r_c$ is given as solution of the equation:
\begin{eqnarray}
\label{ref}
L\left(X_{e}\right)=\frac{1}{2}\pi r^{2}_{c}k_{L}F_{L}\left(X_{t}\right)
\end{eqnarray}
which links also $r_e$ to the cut-off radius $R_t=r_t=a_t$.

Referring to DM distribution its power-law is given by Eq.(\ref{dmdp}) so that in normalized form it becomes of this kind:
\begin{equation}
\label{dmp}
f_D(\xi_D)=\frac{2}{1+(C_D\xi_D)^{d}}
\end{equation}
where, $\xi_D=r/a_D$, $C_D=a_D/r_{oD}$ is the DM concentration, $d\ <\ 1$ and the normalization is done to the
density value $\rho_o(\xi_D=1/C_D)$ at the scale radius $r_{oD}$.
As soon as the concentration of the King's component is fixed for the whole galaxy family\footnote{According to SB9, if we assume: $r_t/r_c=10$ that is the B-component concentration, $C_B=10$, then it turns to be: $r_e/r_c=1.70$.}, then $a_t$ turns to be linearly proportional to $r_e$. Now the next step regards the key point that is to connect $a_D$ of  Eq.($\ref{at})$ to the DM halo mass by cosmology. 


\section{Cosmological framework}

It is worth recalling that the key point to explain the $Mass-Radius$ relationship comes out from the cosmology.  If otherwise, neither all the well known scaling relationships for galaxies nor the FP, which implies their existence,  could be interpreted (see, LS5, SB9). That was one of our first and most important consequences of CVMT, in spite of the apparent paradox that the FP, as a whole, does contain a degeneracy in respect to the initial density perturbation spectrum (first pointed out by Djorgovski, 1992).
 
For a scale-free power spectrum, $P(\kappa)\sim \kappa^n$, the mass variance of DM becomes: 
\begin{equation}
\sigma^2_{M_D}(t)\sim D^2(t)M_D^{-(n+3)/3}
\end{equation}
where the {\it self-similar} linear growing factor, $D(t)$, allow us to describe the general density perturbation as: $\delta(\vec{x},t)=\delta(\vec{x}) D(t)$.

\subsection{$\Lambda$CDM scenario}

We will refer to the cosmological standard model derived from {\it WMAP} precision data only (Spergel et al., 2003; Spergel et al., 2007; Binney \& Tremaine, Chapt.9, 2008) for a flat $\Lambda$-dominated universe: i.e., $\Lambda$CDM model defined by the following parameters: $\sigma_8=0.9\pm 0.1; h=0.72\pm0.05$; matter density, $\Omega_mh^2= 0.14\pm 0.02$, baryon density, $\Omega_bh^2=0.024\pm0.001$ and primordial spectal index $n_s=0.99\pm0.04$ which is consistent with the Harrison-Zel'dovich scale-invariant value ($n_s=1$).
The trend of mass-variance spectrum $\sigma_{M_D}$ at equivalence between matter and radiation (after that epoch the microphysics may never affect the DM fluctuations) is shown in Fig. 1 as function of $M_D$.
\begin{figure}
\resizebox{\hsize}{!}
{\includegraphics{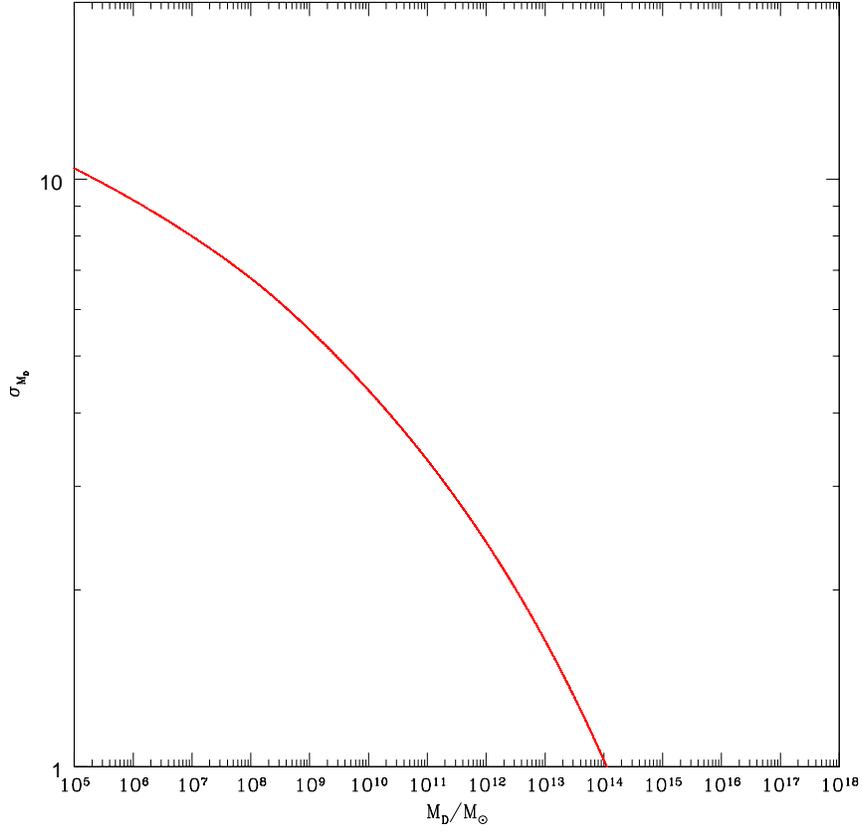}}
\caption{The trend of mass variance $\sigma_{M_D}$, at equivalence epoch, as function of $M_D$, for a flat $\Lambda-$dominated universe derived from {\it WMAP} precision data ($\Lambda$CDM model defined by the following parameters: $\sigma_8=0.9\pm 0.1; h=0.72\pm0.05; \Omega_mh^2= 0.14\pm 0.02; \Omega_bh^2=0.024\pm0.001$;  $n_s=0.99\pm0.04$; see text).}
\label{fig:lcdm}
\end{figure}

In this {\it self-similar model}, an {\it effective} final spectrum index, $n_{eff}=n_e$, may be obtained by local slope of $\sigma_{M_D}$  at a given $M_D$ value, as follows:
\begin{equation}
\label{locs}
\alpha_{eq}=-\frac{{\rm\,d}\log\sigma_{M_D}(t_{eq})}{{\rm\,d}\log M_D}=(n_e+3)/6
\end{equation}
At the halo virialization the following scaling laws hold (see, SB9): 
\begin{eqnarray}
\label{adM}
a_D\sim M_D^{1/\gamma'};\\
\rho_{oD}\sim M_D^{-(n_e+3)/2}\\
\label{gammap}
1/\gamma'=\frac{(5+n_e)}{6}=\frac{3\alpha_{eq} +1}{3}
\end{eqnarray}
where $a_D$ is the halo virial radius and $\rho_{oD}=\rho(r_{oD})$ according to an assumed Zhao's (1996) density profile.

\begin{figure}
\resizebox{\hsize}{!}{\includegraphics{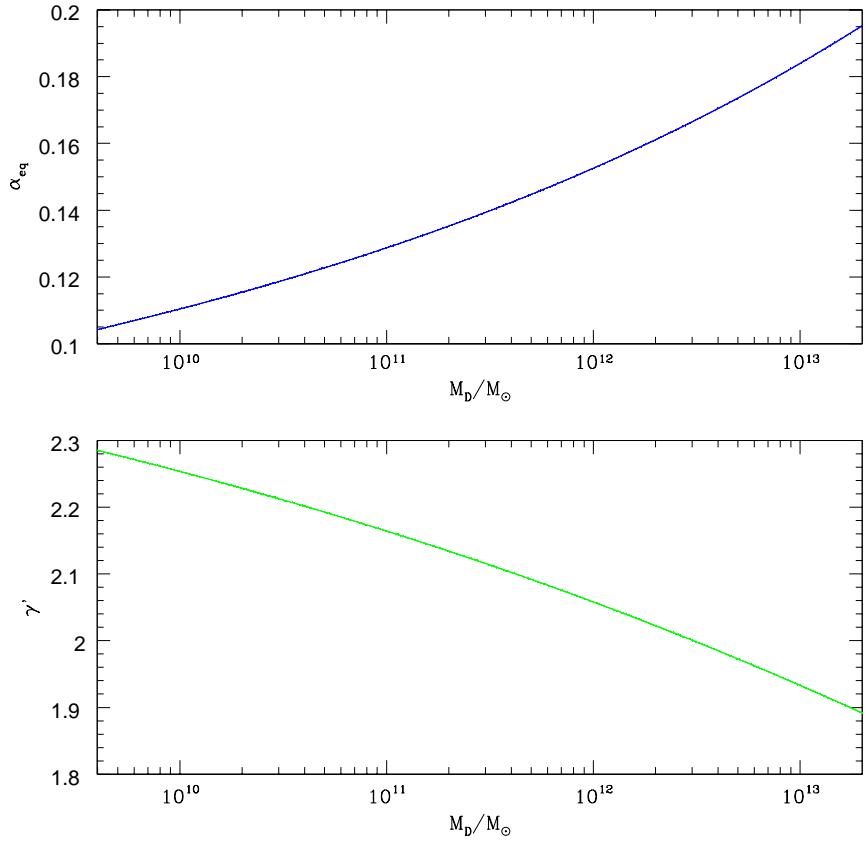}}
\caption{The trend of the local slope, $\alpha_{eq}$, Eq.\,(\ref{locs})
(top panel) and of $\gamma'$, Eq.\,(\ref{gammap}) (bottom panel) as function of DM halo mass in the mass range here considered.}
\label{fig:lcdm_panel}
\end{figure}

\subsection{Scaling laws}
 
Coming back to Eq.(\ref{at}) the scaling law of $r_e$ with $M_B$ for the whole homology family reads: 
\begin{equation}
r_{e}\sim a_t \sim \left(\frac{M_{B}}{M_{D}}\right)^{\frac{1}{3-d}}a_{D}\sim\frac{M^{\frac{1}{3-d}}_{B}}{m^{\frac{1}{3-d}}M^{\frac{1}{3-d}}_{B}}m^{\frac{1}{\gamma'}}M^{\frac{1}{\gamma'}}_{B}
\end{equation}
owing to (\ref{adM}), by definition: $M_D=m M_B$ and under the assumption that bright and dark matter profiles are universal ($\nu_V'$ and $\nu_{\Omega B}$ are depending on both and inner profiles, respectively).
It is to be underlined that in the CVMT all the three main quantities of FP: $r_e$, $I_e$, $\sigma_o$, may be expressed in terms of only two independent parameters, as Brosche as pioneer discovered (Brosche, 1973), e.g. $m$ and $M_B$ (LS5, SB9), and have to scale with them by means of exponents in which enter only two quantities: $d$, i.e., the densiy profile of DM and $\gamma'$ related to the cosmological local slope of mass variance. Indeed the following nice scaling laws have to hold (LS5, SB9): 
\begin{eqnarray}
\label{re}
r_e\sim m^rM_B^R~;~~r=\frac{(3-d)-\gamma'}{\gamma'(3-d)};~R=1/\gamma'\\
\label{ie}
I_e\sim m^i M_B^I~;~~I=i=2\frac{\gamma'-(3-d)}{\gamma'(3-d)}=-2r\\
\label{sig}
\sigma_o\sim m^sM^S_B~; s=-\frac{1}{2}\frac{(3-d)-\gamma'}{\gamma'(3-d)};~S=\frac{1}{2}
\frac{\gamma'-1}{\gamma'}
\end{eqnarray}
To be noted that the bright density distribution doesn't enter. 


\section{Theoretical fits to observations}

The best pioneeristic attempts to find a relationship $M_*-r_e$ appear in the literature due to Chiosi et al. (1998) and Tantalo et al. (1998) starting from the data of Carollo et al. (1993) (see, Contini, 2008). It reads as follows: 
\begin{equation}
\label{tantalochiosi}
r_e = 17.13~ (M_{B,12})^{0.557}
\end{equation}
where $r_e$ is in kpc and $M_{B,12}=M_B/(10^{12} \,M_{\odot}) \,$. The data refer to 42 ETGs, 11 S0 + 1 which probably could be a spiral. The \textit{fit} is given in Tantalo et al. (1998). 
As we will see, relationships of kind (\ref{tantalochiosi}) may gain a full explanation inside the framework of CVMT. 
We refer here to effective radii and star subsystem masses determined in a recent attempt  
(Tal09) for local elliptical galaxies, using model populations and a large uniform data set from a well populated homogeneous sample
( Prugniel \& Simien (1996 (PS96); 1997 (PS97)).  More specifically, Tal09 aim to carefully consider all the factors which may enter the FP. 
Many accurate, interesting fits and mass models are given therein.

Better and more extended
samples of ETGs are available now in the literature, in particular the data from
SDSS, where  a similar {\it mass-size} relation is shown (e.g., Bernardi et al. 2003)
while mass determination is unnecessary in testing the Faber-Jackson relation
(e.g.,  Nigoche-Netro et al. 2010).   For this reason, as a first step
Tal09 results shall be used in the following.

The related {\it mass-size} relation (Tal09, Fig.\,4) when binned averaged points are considered, has a best fitting mean slope over the full range of the data, equal to: $0.58\pm0.05$, in agreement with previous estimates of typically $\sim0.6$ (Bernardi et al. 2003; Shen et al. 2003; Mamon \& Locas 2005; Napolitano et al. 2005)
and recent estimates of typically $\sim0.55$ (Bernardi et al., 2011, hereafter Bern11).
Moreover two mass  regimes are identified: faint galaxies ($M_*< 10^{11.1}M_{\odot})$ with a slope: $0.36\pm0.13$, and bright galaxies ($M_*\geq 10^{11.1}M_{\odot}$) with a greater slope: $0.73\pm0.12$. At fixed mass ratio $m$, this trend appears immediately in agreement with the theoretical trend of $1/\gamma'$ expected by the scaling relation ($\ref{re}$) where the exponent increases at increasing $M_D$ because the local slope of the cosmological mass variance increases (Eqs. (\ref{locs}), (\ref{gammap})). 

\begin{figure}
\resizebox{\hsize}{!}{\includegraphics{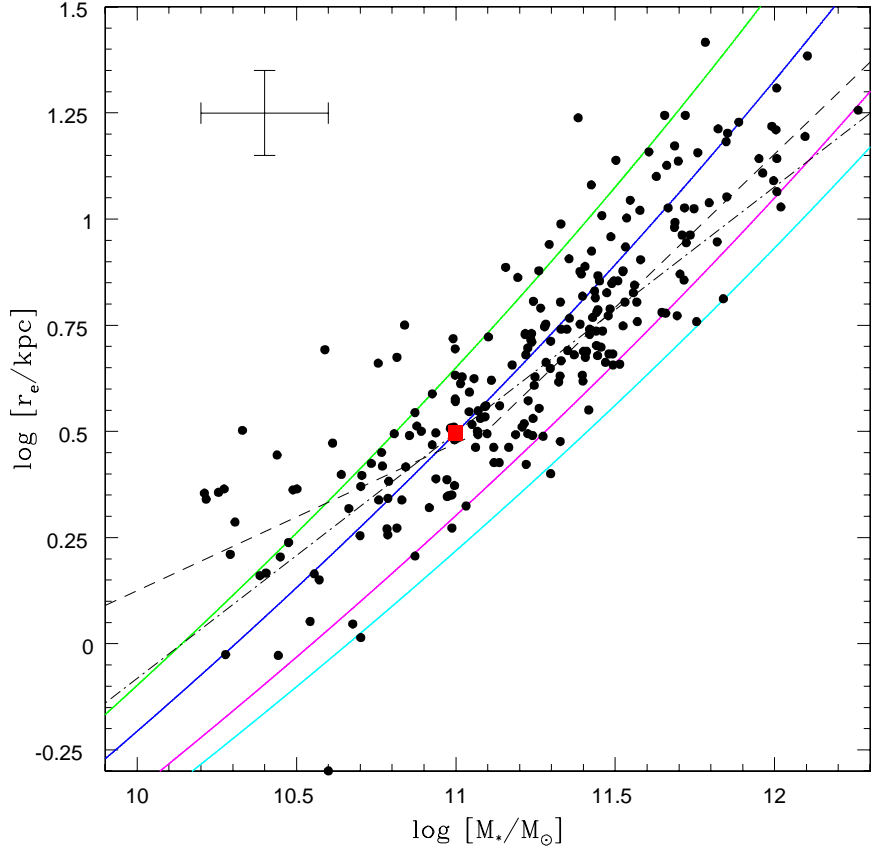}}
\caption{{\it Mass-Radius} data  of local ETGs from PS96 (see, Tal09). The solid lines in colours are the theoretical curves given by Eq.(\ref{tfits}) when the mass ratio $m$ assumes (bottom-up) the values: $1,4,10$, and $d=0.5$. The red square refers to the calibration point for which it is assumed $m^o=4$ so that the other corresponding parameters entering (\ref{tfits}), are: ($M_B^o= 1\cdot10^{11}M_{\odot},\ r_e^o=3.133\ kpc,\ r_o=0.076,\ \gamma'_o=2.103$). The
bended dashed line is the fit given by Tal09, corresponding to the two regimes of low and high masses. The dot-dashed straight line shows the mean mass-size relation found by the same authors. On the top left corner the rms errors are shown as a cross. The limit case $m=0.5\simeq m_l$ is also given as lowest curve delimiting the underlying avoidance region. See the on-line version for colours.}
\label{mo4_d05}
\end{figure}

More precisely the CVMT is able to differentiate the contribution of either mass ratio $m$ or $M_B$ (see, (\ref{re})), 
both having a variable exponents in order to produce a continuous set of curves on the data plane ($log\ M_*, log\ r_{e}$) (Tal09, Fig.4). That allows us to improve the resolution of the relationship we are looking for, taking into account the trends of the points which show, in Tal09's data, the individual galaxies of the sample.
According to the theoretical scaling law (\ref{re}), we have:
\begin{eqnarray}
\label{tfits}
~~~~~~~~~~~~~~r_{e}=r^{o}_{e}\frac{m^{r}}{\left(m^{o}\right)^{r_{o}}}\frac{M^{\frac{1}{\gamma'}}_{B}}{\left(M^{o}_{B}\right)^{\frac{1}{\gamma'_{o}}}}=\ K m^{r}M_{B}^{\frac{1}{\gamma'}}\\
\label{reMm}
K=\ r^{o}_{e}\frac{1}{\left(m^{o}\right)^{r_{o}}\left(M^{o}_{B}\right)^{\frac{1}{\gamma'_{o}}}}= const.
\end{eqnarray}
The values: ($M^{o}_{B}$, $r^{o}_{e}$, $r_{o}$, $\gamma'_{o}$) refer to an arbitrary calibration galaxy chosen to normalize the trends in the sample.

\subsection{Calibration}

Because in Tal09 $M_B^o\simeq 10^{11} M_{\odot}$ works as characteristic mass scale between the two mass regimes of {\it faint} and {\it bright} subsamples, we select it as mass calibration point. The corresponding effective radius is $r_e^o=3.133 kpc$ which has to read on the mean mass-size relation found by Tal09 with the mean slope $\alpha_M=0.58\pm0.05$. About the mass ratio $m^o$, there is not an {\it a priori} value to be chosen then it will be considered a free parameter even if some hints may come from cosmology. It is to be underlined that the values of both $r_o, \gamma'_o$ are determined by $m^o$ because it settles the amount of dark matter halo which in turn gives the local slope (\ref{locs}) in the cosmological mass variance $\sigma_{M_D}$.    
 
According to the mean $68\%$ confidence range given by Spergel et al. (2003) for matter density and baryon density in their $\Lambda$CDM model, we obtain a corresponding range for the cosmological ratio: $M_D/M_B$ as:
\begin{equation}
\label{cosm}
m=\frac{1-\frac{\Omega_bh^2}{\Omega_mh^2}}{\frac{\Omega_bh^2}{\Omega_mh^2}};\ \ 3.8\leq m \leq 5.9
\end{equation}
As scouting sake a more extended range ($m=1\div 10$) will be considered assuming as reference value: $m=4$. Moreover, according with the CVMT we know that an inferior $m-$limit has to exist, given by Eq.(\ref{ml}), which for the cases here considered turns to be in the range: $1\div 0.5$ (Bindoni, 2008). For sake of investigation the range assumed will be: $m=1\div 10$. For special aims we will consider also the extreme limit $m=0.5\simeq m_l$.

\subsection{Theoretical relationships}

Three {\it mass-size} theoretical relationships of kind (\ref{tfits}) may be performed by assuming different values of $m^o$ for the calibration point in each of them (e.g., $m^{o}$=j=1, 4, 10). Starting from each of these curves with label $j$ Eq.(\ref{tfits}) yields a {\it continuum} of curves by changing $m$. For sake of simplicity we consider only a couple of them for each $j$: $m$=4, 10; $m$=1, 10; $m$=1, 4, respectively. In Fig. 3 the triplet of curves corresponding to: $m^o$=j=4; $m$=1, 10 is shown.
They are drawn at fixed $m$ but with the exponents $r$ and $1/\gamma'$ changing with $M_B$.
The constants (\ref{reMm}) are:
\begin{equation}
\label{kmi}
K^j=\ K^j( r^{o}_{e}, m^{o}=j, M^{o}_{B}, r_{o}, \gamma'_{o});\ \ j=1,4,10 
\end{equation}
The equations of the corresponding curves become ($i$=1, 4, 10): 
\begin{equation}
\label{thecuves}
r_e=
\begin{cases}
 K^1\ (m_i^j)^{r}\ M_{B}^{\frac{1}{\gamma'}}\ ; m^o=j=1;\ \ m_i^1=\ 1,\ 4,\ 10& \cr 
 K^4\ (m_i^j)^{r}\ M_{B}^{\frac{1}{\gamma'}}\ ; m^o=j=4;\ \ m_i^4=\ 1,\ 4,\ 10& \cr
 K^{10}\ (m_i^{j})^{r}\ M_{B}^{\frac{1}{\gamma'}}\ ; m^o=j=10;\ m_i^{10}=\ 1,\ 4,\ 10&  \cr
\end{cases}
\end{equation}
The common property of all the curves is to have a positive local slope increasing with $m$ at fixed value of $M_B$. The triplet of curves, shifts towards bottom as soon as the value of $m^o$ attributed to the calibration point increases from $1$ to $10$. The case $m^o=4$ seems to produce a better fit to the observed spread  which remains greater than the typical vertical rms error bar shown in Fig.\,\ref{mo4_d05}.

\subsection{Linearization}

Our aim is to linearize the theoretical curves (\ref{thecuves}) in order to obtain useful numerical relationships.
In logarithmic form the set of equations becomes:
\begin{equation}
y_i^j(x)=\tilde{K}^j+\xi_i^j(x)\ \tilde{m}_i^j+\zeta_i^j(x)\ x;\ (i,j=1,4,10)
\end{equation}
with the following definitions:
\begin{equation}
\label{defin}
\begin{cases}
y=log\ r_e& \cr 
x=log\ M_B& \cr
\xi_i^j(x)= r(M_B)& \cr
\zeta_i^j(x)=\frac{1}{\gamma'(M_B)}& \cr
\tilde{m}_i^j= log\ m_i^j& \cr
\tilde{K}^j= log\ K^j& \cr
\end{cases}
\end{equation}
They may be linearized by fixing a reference point in each curve with coordinates \footnote{In our curves the x-coordinate always corresponds to $\bar{x}_i^j=\bar{x}=11.00$.} ($\bar{x}_i^j, \bar{y}_i^j)$ and taking  the following local slope at the same point :
\begin{equation}
\label{rslope}
p_i^j(\bar{x})=\left(\frac{dy^j}{dx}\right)_{\bar{x}}=\left(\frac{d\xi_i^j}{dx}\right)_{\bar{x}}\tilde{m}_i^j+\left(\frac{d\zeta_i^j}{dx}\right)_{\bar{x}}\ \bar{x}+ \zeta_i^j(\bar{x})
\end{equation}
It is noteworthy that the theoretical slope at a reference point $\bar{x}$ preserves memory of the cosmological mass variance either through its first derivative (see, Eqs. (\ref{locs}, \ref{gammap})), inside the third contribution ($\zeta$ term in Eq.(\ref{rslope})) or its second derivative (first and second term of the same equation).
The transformed linear equations of set (\ref{thecuves}) become:
\begin{equation}
\label{lineareq}
z_i^j(x)=\tilde{K'}^j_i+ p_i^j(\bar{x})\ x\ \rightarrow\ r_e=K'^j_i\ M_B^{p_i^j}
\end{equation}
where:
\begin{eqnarray}
\tilde{K'}^j_i= \bar{y}_i^j-p_i^j(\bar{x})\ \bar{x}\\
\bar{y}_i^j=z_i^j(\bar{x})
\end{eqnarray}
It may be interesting to distinguish between the two contributions: one due to the derivative of $1/\gamma'(M_B)$ and the other one  to the derivative of $r(M_B)$. Both added to the term $\zeta_i^j(\bar{x})$ enter the slopes $p_i^j(\bar{x})$. To this aim we define:
\begin{eqnarray}
q_i^j(\bar{x})=\left(\frac{d\xi_i^j}{dx}\right)_{\bar{x}}\tilde{m}_i^j\\
g_i^j(\bar{x})=\left(\frac{d\zeta_i^j}{dx}\right)_{\bar{x}}\ \bar{x}\\
\label{tslop}
p_i^j(\bar{x})=q_i^j(\bar{x})+g_i^j(\bar{x})+\zeta_i^j(\bar{x})
\end{eqnarray}
The different contributions are evaluated in Tab.4.
It is noteworthy that the trend of $\zeta(x)=\frac{1}{\gamma'(M_B)}$, which determines both the slopes $q(x)$, $g(x)$ and in turn $p(x)$, is depending on the value of $M_D$. That means: at fixed $M_B$, the slopes have to be the same if $m$ is the same. So curves with the same label $i$ and different $j$ are simply the same curves vertically translated one in respect to the other owing to the different values of $log\ r_e$. Furthermore, at fixed $j$, a curve at higher $m$ is simply obliqually shifted over that at lower $m$, towards higher values of $M_B$. Moreover, at fixed $d$, whichever it is, by definition of (\ref{re}), it occurs:
\begin{equation}
\label{egde}
\left(\frac{d\xi_i^j}{dx}\right)_{\bar{x}}=\left(\frac{d\zeta_i^j}{dx}\right)_{\bar{x}}
\end{equation}

That allows us to collect in a unique Tab.1 the trends of $\gamma'$, $d\zeta/dx$ and $p$ in the three cases $m=1,4,10$ instead of handling with three tables.

\begin{table}
\caption{The trends of $\gamma'$ and the first derivative  of $\zeta=1/\gamma'$ as function of the baryonic mass $M_B$ and $m$ are shown. The end product $p$, according to (\ref{tslop}), is also given for any $d$ (see, Figs.3,5).} 
\centering
\begin{tabular}{|c|c|c|c|c|c|c|c|c|c|}
\hline
\hline
$log\ M_B$ & $\gamma'(x)$ & $\gamma'(x)$ & $\gamma'(x)$ & $d\zeta(x)/dx$ & $d\zeta(x)/dx$ & $d\zeta(x)/dx $ & $p(x)$ & $p(x)$ & $p(x)$ \\
$x$ & m=1 & m=4 & m=10 & m=1 & m=4 & m=10 & m=1 & m=4 & m=10 \\
\hline         
9.9 & 2.262 & 2.210 & 2.174 &&&&&& \\
10.0 & & & & 0.016 & 0.018 & 0.020 & 0.603 & 0.650 & 0.687 \\
10.1 & 2.245 & 2.192 & 2.155 &&&&&& \\
10.2 & & & & 0.017 & 0.019 & 0.022 & 0.623 & 0.668 & 0.715 \\
10.3 & 2.228 & 2.174 & 2.134 &&&&&& \\
10.4 & & & & 0.018 & 0.021 & 0.023 & 0.636 & 0.694 & 0.735 \\
10.5 & 2.211 & 2.154 & 2.113 &&&&&& \\
10.6 & & & & 0.019 & 0.022 & 0.025 & 0.653 & 0.712 & 0.763 \\
10.7 & 2.193 & 2.134 & 2.091 &&&&&& \\
10.8 & & & & 0.020 & 0.023 & 0.025 & 0.671 & 0.739 & 0.781 \\
10.9 & 2.174 & 2.113 & 2.069 &&&&&& \\
11.0 & & & & 0.020 & 0.024 & 0.027 & 0.687 & 0.755 & 0.808 \\
11.1 & 2.155 & 2.091 & 2.047 &&&&&& \\
11.2 & & & & 0.022 & 0.025 & 0.029 & 0.715 & 0.781 & 0.843 \\
11.3 & 2.134 & 2.069 & 2.023 &&&&&& \\
11.4 & & & & 0.023 & 0.027 & 0.031 & 0.735 & 0.813 & 0.879 \\
11.5 & 2.113 & 2.046 & 1.998 &&&&&& \\ 
11.6 & & & & 0.025 & 0.029 & 0.033 & 0.763 & 0.847 & 0.915 \\
11.7 & 2.091 & 2.022 & 1.972 &&&&&& \\ 
11.8 & & & & 0.025 & 0.031 & 0.034 & 0.781 & 0.880 & 0.941 \\  
11.9 & 2.069 & 1.997 & 1.946 &&&&&& \\ 
12.0 & & & & 0.027 & 0.032 & 0.036 & 0.808 & 0.905 & 0.980 \\ 
12.1 & 2.047 & 1.972 & 1.920 &&&&&& \\
12.2 & & & & 0.029 & 0.034 & 0.038 & 0.843 & 0.941 & 1.028 \\ 
12.3 & 2.023 & 1.946 & 1.892 &&&&&& \\   
\hline
\end{tabular}
\label{tab:const1}
\end{table}


\section{Useful Relationships}

According to the linearization procedure performed in subsect.\,4.3, we may relax the condition of variability of $r$ and $\gamma'$ along the $M_B$ coordinate, as occurs in the theoretical Eqs.(\ref{thecuves}), and consider the three straight lines passing through the calibration point having, at this point, the same slopes of the three curves of the set (\ref{thecuves}) with: $j=1,m=1; j=4, m=4, j=10, m=10$.
The linear equations, as approximations of Eqs.(\ref{thecuves}), are, respectively:
\begin{equation}
\label{uslirp}
\frac{r_e}{\rm kpc}=
\begin{cases}
15.240\ (M_{B,12})^{0.687};\ \ m=1& \cr 
17.824\ (M_{B,12})^{0.755};\ \ m=4& \cr
20.137\ (M_{B,12})^{0.808};\ \ m=10& \cr
\end{cases}
\end{equation}
where $M_{B,12}=M_{B}/(10^{12}\ M_{\odot})$ and $d=0.5$. It is to be noted that the second equation is not too far from the preliminary pioneeristic one found by Tantalo et al. (1998) using the data of Carollo et al. (1993).
In Tab. 2 the scatters between these linear approximations and the right theoretical curves are summarized.
Then the method may be generalized thanks to the remarkable note that the slope, at fixed $M_B$, is depending only by $m$. So the following three general linear relationships may be built up: 
\begin{equation}
\label{uslinear}
log\ r_e=
\begin{cases}
[\bar{y}- p_1(\bar{x}) (\bar{x})] + p_1(\bar{x})\ log M_B;\ \ m=1& \cr 
[\bar{y}- p_4(\bar{x}) (\bar{x})] + p_4(\bar{x})\ log M_B;\ \ m=4& \cr 
[\bar{y}- p_{10}(\bar{x}) \bar{x})] + p_{10}(\bar{x})\ log M_B;\ \ m=10& \cr 
\end{cases}
\end{equation}
which hold whatever is the calibration point ($\bar{x}, \bar{y}$), reading the corresponding slope, at fixed $m$, in Tab.1.
As soon as the calibration point is ($\bar{x}=11.00,\bar{y}= 0.496$) the set (\ref{uslinear}) transforms into that of Eqs.(\ref{uslirp}).

\begin{table}
\caption{Relative scatter between values from the straight lines of Eqs.(\ref{uslirp}) and $r_e^*$ from the theoretical curves of Eqs.(\ref{thecuves}), at fixed coordinate $log\,M_B$ and mass ratio $m$.}
\centering
\begin{tabular}{|c|c|c|c|c|}
\hline
\hline
$log\ M_B$ & m & $r_e^*(kpc)$ & $r_e$(kpc)& $\frac{\Delta r_e}{r_e}\cdot\%$ \\
\hline         
9.90 & 1 & 0.617 & 0.550 & 11 \\
12.30 & 1 & 31.189 & 24.493 & 21 \\
9.90 & 4 & 0.531 & 0.463 & 13 \\
12.30 & 4 & 40.137 & 30.023 & 25 \\
9.90 & 10 & 0.476 & 0.405 & 15 \\
12.30 & 10 & 49.333 & 35.184 & 29 \\
\hline
\end{tabular}
\label{tab:const2}
\end{table}

\begin{table}
\caption{Mean theoretical slope values as function of $M_B$ (column (1)) at lower ($M_B<10^{11}$ $M_{\odot}$) and higher ($M_B\ge 10^{11}$ $M_{\odot}$) mass regimes (index $l$ and $h$, respectively), in the cases: $m=\ 1,\ 4,\ 10$ (columns: (2), (3), (4)) are shown. The respective mean theoretical slopes in the whole range are given in columns: (5), (6), (7). Columns (8), (9), refer those of Tal09's for comparison.}
\centering
\begin{tabular}{|c|c|c|c|c|c|c|c|c|}
\hline
\hline
$log\ M_B$ & $\frac{(p_1)_l}{(p_1)_h}$ & $\frac{(p_4)_l}{(p_4)_h}$ & $\frac{(p_{10})_l}{(p_{10})_h}$ & $\bar{p}_1$ & $\bar{p}_4$ & $\bar{p}_{10}$ & $\frac{p^T_l}{p_h^T}$ & $\bar{p}^T$ \\
(1) & (2) & (3) & (4) & (5) & (6) & (7) & (8) & (9) \\ 
\hline
9.90& & & & & &	& & \\
 & & & & & & & & \\
 & 0.637 & 0.693 & 0.736 & & & & $0.36\pm0.13$ & \\
 & & & & & & & & \\
11.00 & & & & 0.700 & 0.769 & 0.825 & & $0.58\pm0.05$ \\
 & & & & & & & & \\
 & 0.762 & 0.846 & 0.913 & & & & $0.73\pm0.12$ & \\
 & & & & & & & & \\
12.30 & & & & & & & & \\
\hline
\end{tabular}
\label{tab:const3}
\end{table}

\begin{table}
\caption{Values related to calibration red point: $M_B^o$=$1.00\cdot 10^{11}$ $M_{\odot}$, $r_e^{o}$=3.133 kpc, entering (\ref{tfits}) and (\ref{thecuves}). The corresponding values of constants $K^j$ in (\ref{kmi}) and (\ref{thecuves}) with $M_B$ in $M_{\odot}$ and $K'^j_i$, which enter the numerical relationships (\ref{lineareq}) and (\ref{uslirp}), are also given in the case $d$=0.5 when $M_B$ is in units of $10^{12}$ $M_ {\odot}$.} 
\centering
\begin{tabular}{|c|c|c|c|c|c|c|c|c|c|}
\hline
\hline
$j=m^o$ & $m_i^j$ & $r_o$ & $\gamma'_o$ & $1/\gamma'_o$ & $K^j(\cdot 10^5)$ & $K'^j_i$ & $p_i^j(\bar{x})$ & $q_i^j(\bar{x})$ & $g_i^j(\bar{x})$ \\
\hline
1.0	& 1 &	0.062	&	2.164	& 0.462 & 2.594 & 15.240 & 0.687 & 0.0000 & 0.225 \\
 " & 4 & " & " & " & " & & & & \\
 " &10 & " & " & " & " & & & & \\
4.0	& 1 &	0.076 & 2.103 & 0.476 & 1.637 & & & & \\
 " & 4 & " & " & " & " & 17.824 & 0.755 & 0.0145 & 0.264 \\
 " & 10 & " & " & " & " & & & & \\
10.0 & 1 & 0.086 & 2.058 & 0.486 & 1.159 & & & & \\
 " & 4 & " & " & " & " & & & & \\
 " & 10 & " & " & " & " & 20.137 & 0.808 & 0.0270 & 0.297 \\
\hline
\end{tabular}
\label{tab:const4}
\end{table}


\section{Theoretical slopes vs. observations}

Tab.3 allows us to compare our theoretical mean slopes as function of $M_B$ at different mass ratio $m$, with those given by Tal09 in the  two mass regimes, separated by $M_B= 10^{11}\ M_{\odot}$, which are labelled with the indices: $l$ and $h$ respectively (columns (2),(3),(4) to compare with column (8)). Also the theoretical mean slopes over the total range are given in columns: (5),(6),(7), to be compared with column (9) (Tal09's data). It appears that the mean theoretical slope at high mass regime is in fairly well agreement with the value found by Tal09 if we exclude the case of $m=10$ which leads to a value out of the error bar. According to the $\Lambda$CDM scenario, the typical value expected for the  Dark/Bright mass ratio in galaxy sample considered doesn't be around $10$ but  rather about: $m\simeq 4-5$. Then the theoretical case m=10 is indeed to be ruled out. At lower regimes the mean slope of CVMT is too high. A possible explanation may be the following: the mean dashed straight line of Tal09 goes across the theoretical lines
characterized by a fixed $m$ value (see, Fig.3) so that the fit results flatter. The consequence is that also the mean slopes in the whole range: $\bar{p_1}, \bar{p_4}$, of Tab.3, turns to be out of the maximum limit obtained from observations (column (9) and dot-dashed straight line plotted in Fig.\,\ref{mo4_d05}).


\section{Intrinsic Dispersion vs. DM distribution}

According to the theoretical interpretation given by TCV, the ($M_*-r_e$) relationship is depending on a third parameter, $m$, which enters (\ref{tfits}) with a non-irrelevant exponent $r$. By its definition (\ref{re}) it is manifest that it is a function of $n_e$ via $\gamma'$ and of $d$ which determines the DM distribution. For a typical mass of DM halo on the galaxy scale, the value of $n_e\simeq -2$, then it follows that $\gamma'\simeq 2$ and then $r$ decreases from $r\simeq 1.7$ at $d=0$, towards $r=0$ when $d\rightarrow 1$, reaching the value $r=0.1$ at $d=0.5$. In turn it means that the two exponents of the ($M_*-r_e$) relation are comparable for $d\le0.5$ and then both are to be considered. The  effect of $m$ is to shift the curves by a non-trivial quantity due to the role of $r$.
To a value of $M_*$ in each figure, it has to correspond an intrinsic spread in $r_e$ due to the different possible values of $m$ apart from the measurement rms errors (here assumed to be of $ \pm 20\%$, i.e. $\pm \sigma(log\ r_e) =\pm\sigma_e=0.1$). It means that the same baryonic mass may have DM haloes of different mass values $M_D$.

If we move from a fixed curve of Eq.(\ref{tfits}) (e.g., $m^o=4$) this intrinsic dispersion, at fixed baryonic mass (e.g. $M_B^o=10^{11} M_{\odot}$) and at fixed value of $r_o$ ($d_o=0.5;\gamma'_o=2.103$, Tab.4), is given by:
\begin{equation}
\label{spr}
\left(\frac{\delta r_e}{r_e}\right)_{disp}= r_o\frac{\delta m}{m^o};\ r_o=\frac{(3-d_o)-\gamma'_o}{\gamma'_o(3-d_o)}=0.076
\end{equation}
\begin{figure}
\includegraphics[width=6.5cm]{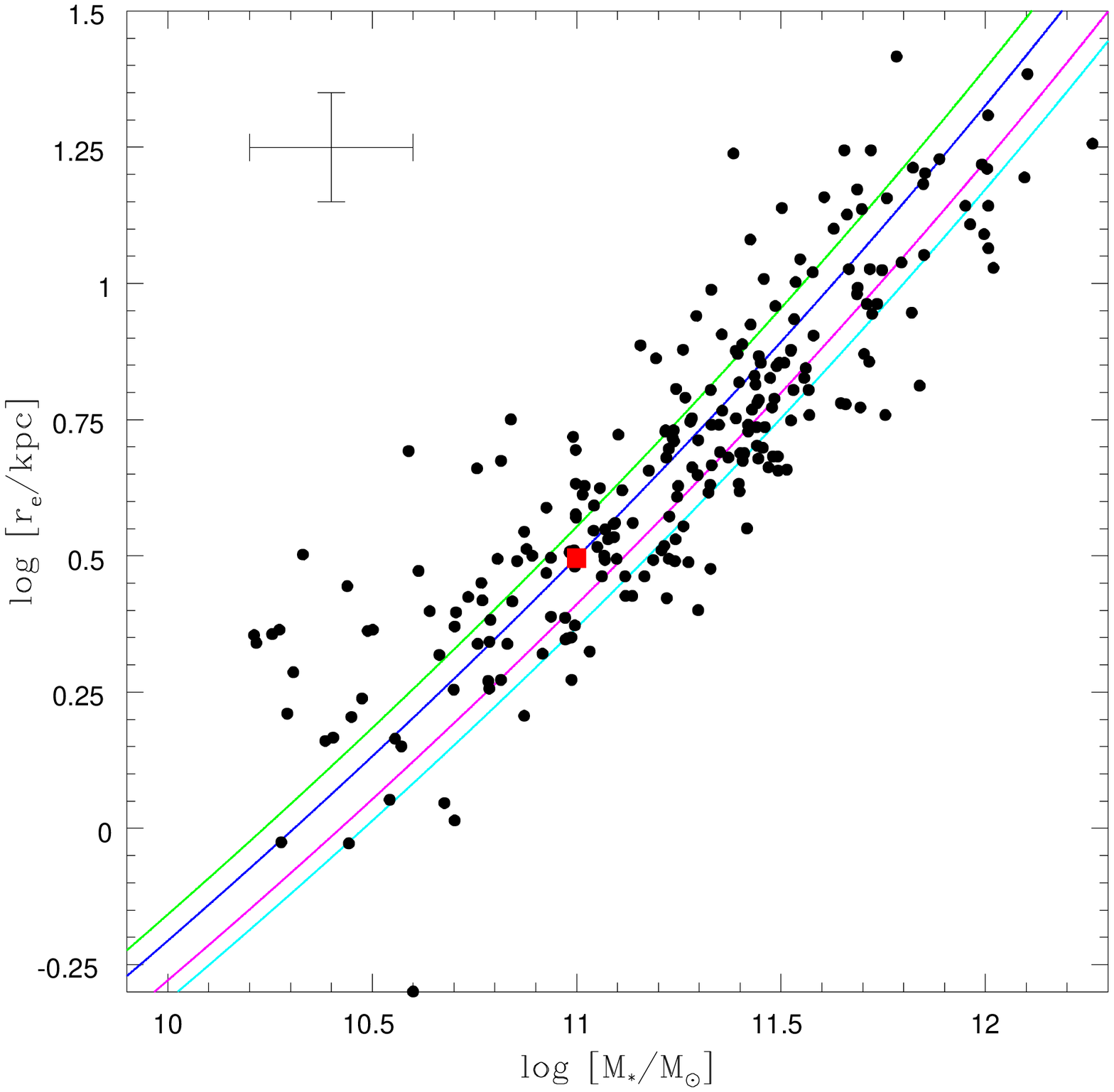}
\includegraphics[width=6.5cm]{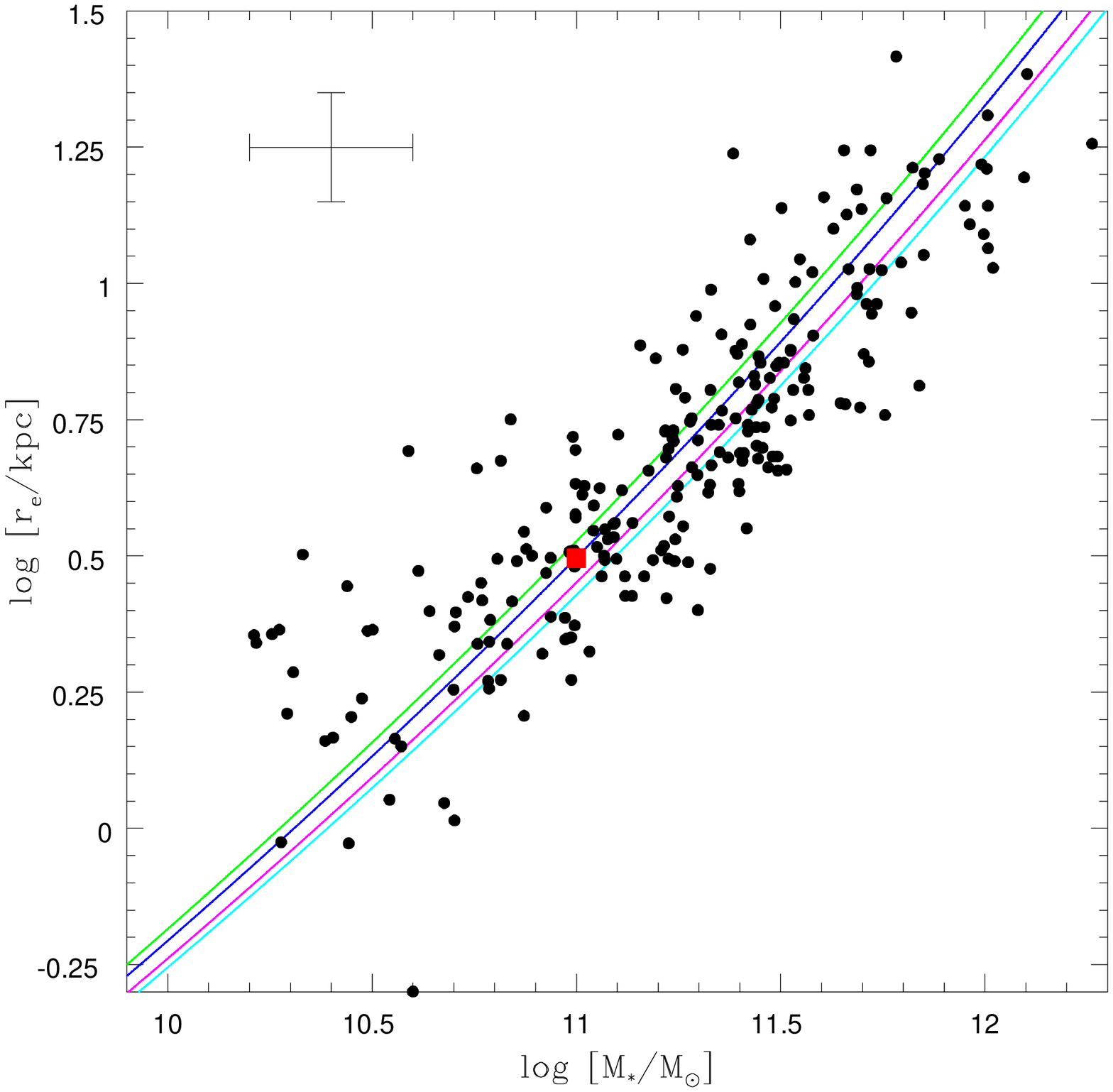}
\includegraphics[width=6.5cm]{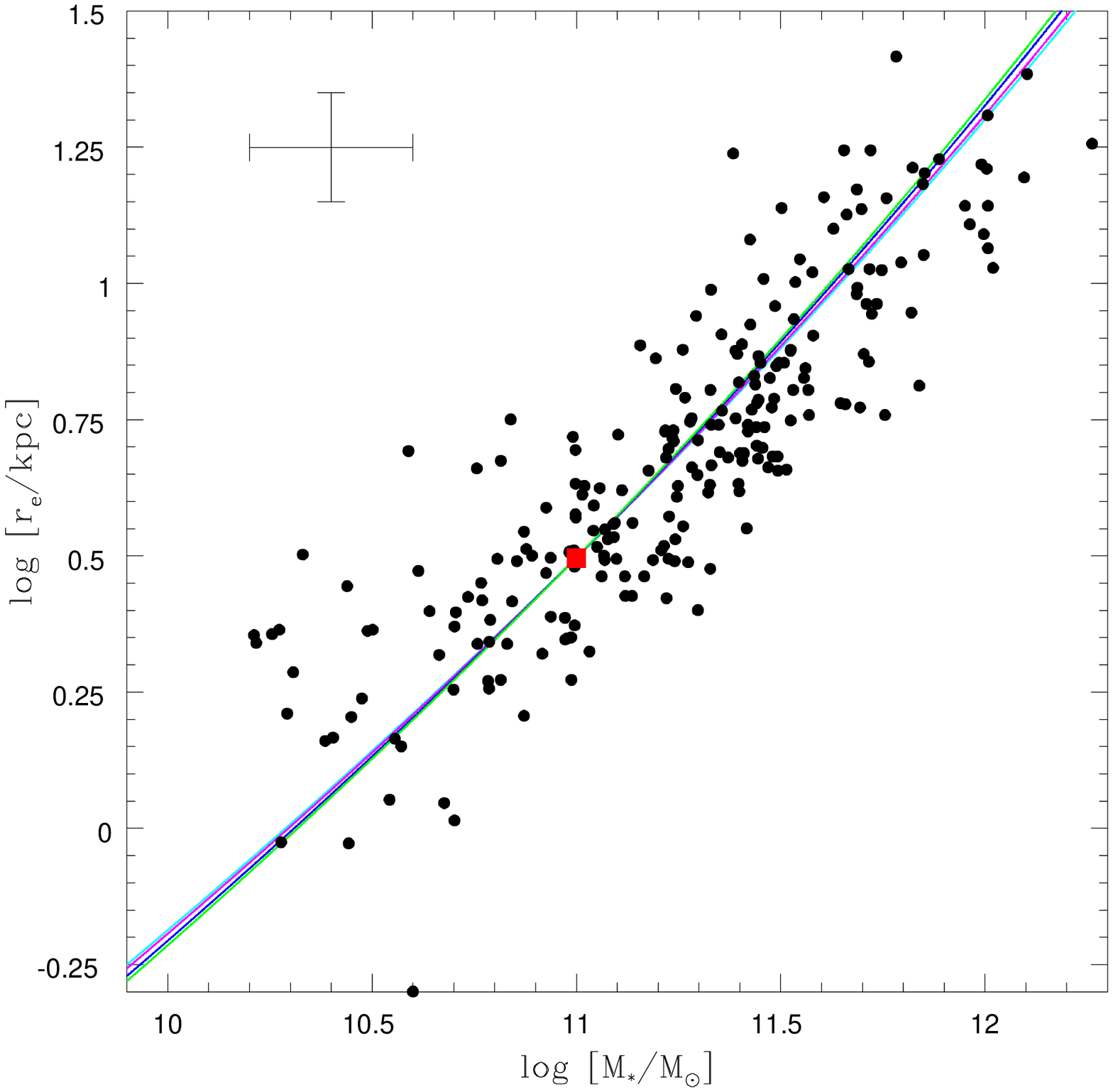}
\caption{Intrinsic dispersion due to the logarithmic slope $d$ of DM distribution in the three cases: $d=0$ (top-left), $d=0.5$ (top-right), $d=0.897$ (bottom-left) by $\gamma'$ distribution frozen to that of the reference curve $m^o=4$. The mean dispersion values at the reference point $M_B=M_B^o$ are given in Tab.5. $r$ becomes 0 for $\gamma'_o=2.103$ when d=0.897 (see, text). See the on-line version for colours.}
\label{<Your label>}
\end{figure}

\begin{figure}
\includegraphics[width=6.5cm]{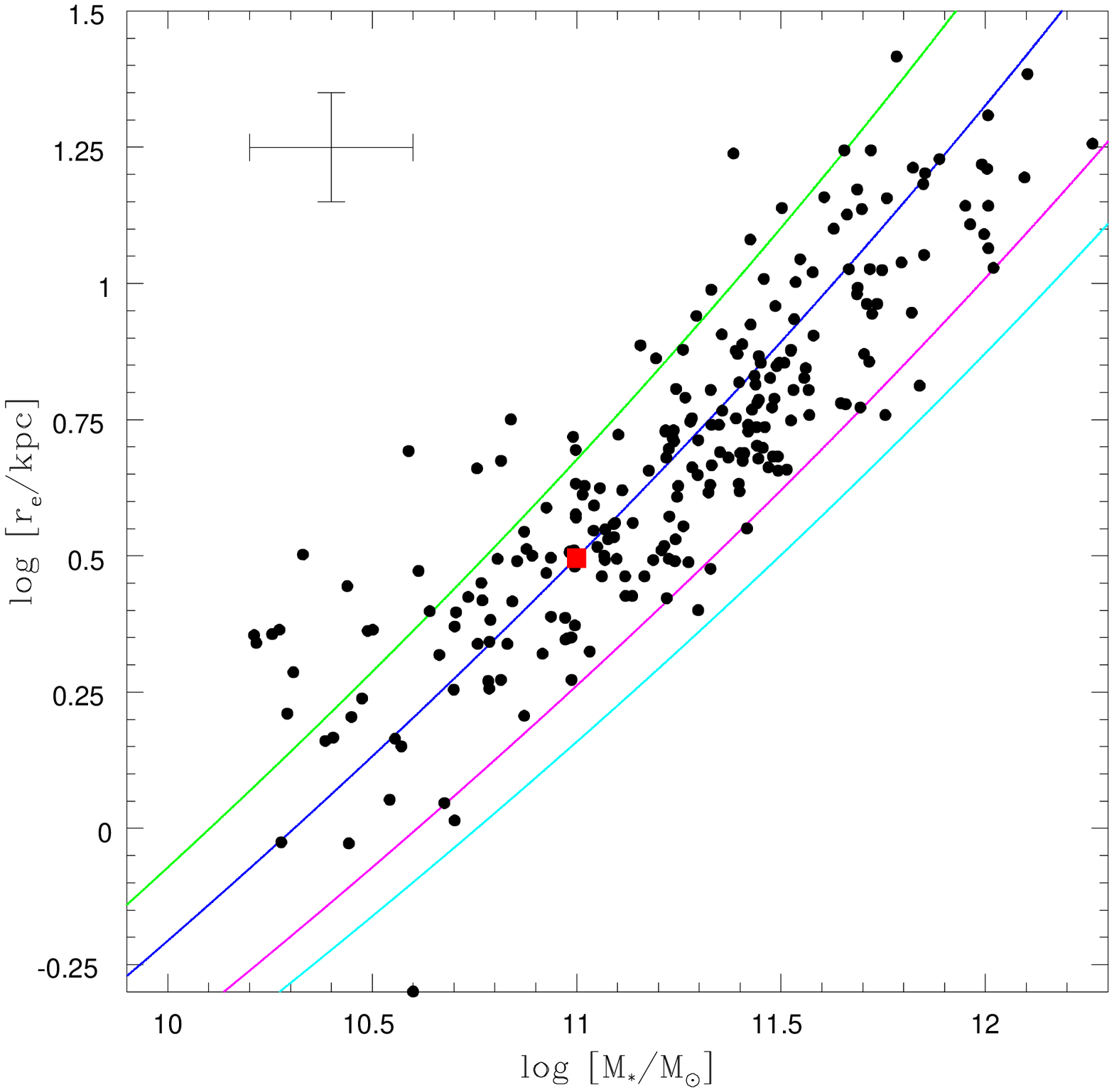}
\includegraphics[width=6.5cm]{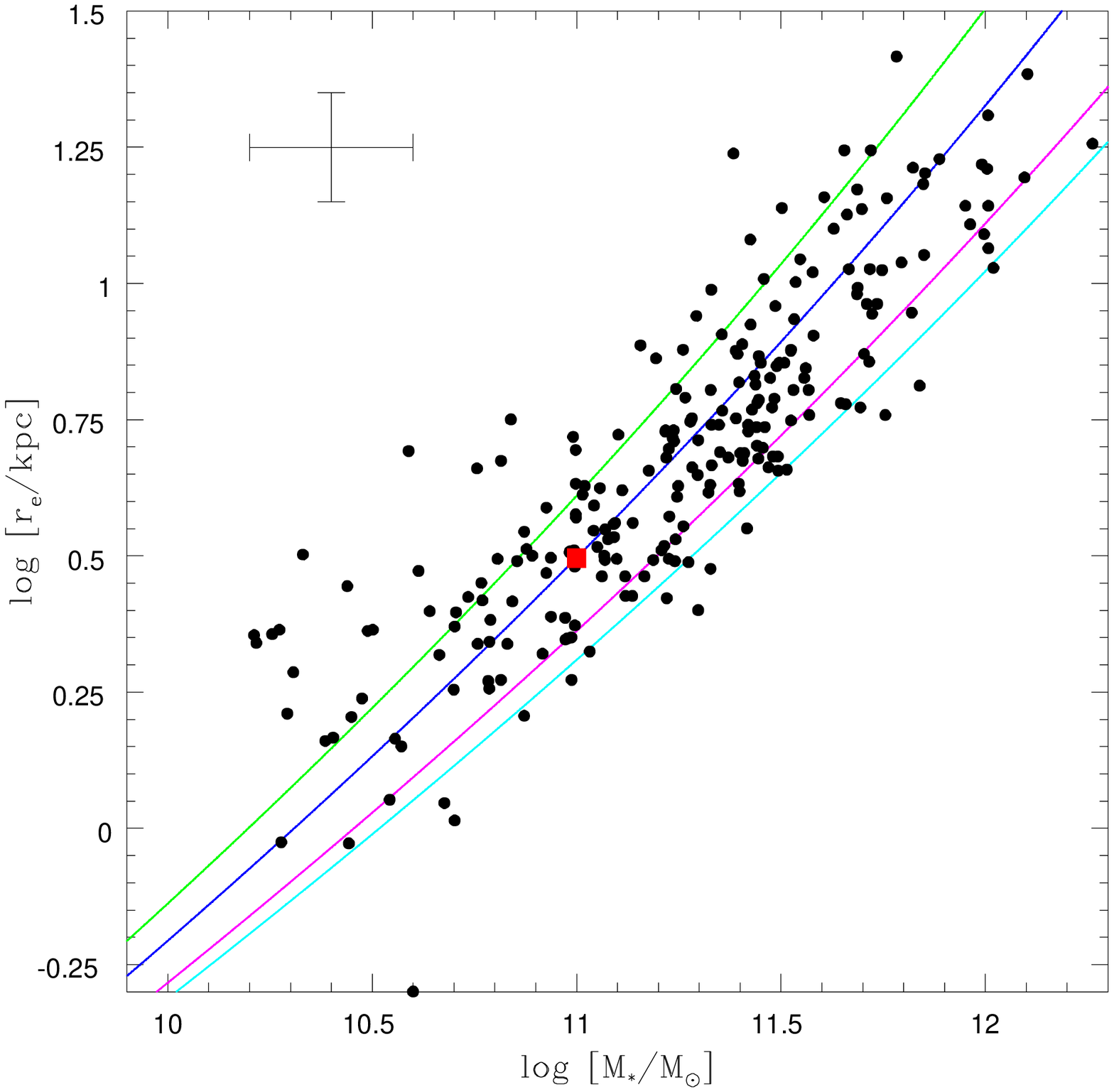}
\caption{As in the previous Fig.3 when the logarithmic slope $d$ of DM density profile becomes $d=0$ (left) and $d=1$ (right) (see, text). To be noted that the curve for $m=m^o=4$ remains the same whichever is the value of $d$; the others shift parallel by changing $d$ if with the same $m$. See the on-line version for colours.}
\label{mo4_d0_d1}
\end{figure}

This scatter, for the same $\gamma'_o$ and the same variation of $m$ depends on the exponent $d$ tuning the DM distribution inside which the bulk of baryonic matter in stars is embedded. Eq.(\ref{spr}) reads as: 
\begin{equation}
\label{sspr}
\left(\frac{\delta r_e}{r_e}\right)_{disp}=-\frac{1}{3-d}\frac{\delta m}{m^o}+ \frac{1}{\gamma'_o}\frac{\delta m}{m^o}
\end{equation}
In Tab.5 the whole wheighted contribution of (\ref{sspr}) as $\Delta log r_e$ is shown in the cases with different $d$ and changing $m$ from $m^o$ {\it up} to $10$ and {\it down} to $1$ together with the {\it mean} dispersion scatter. It should be noted that the intrinsic dispersion becomes comparable with the rms errors as soon as $0\le d\le 0.5$. 
The Eq.(\ref{sspr}) is a differential relationship obtained by deriving the second equation of set (\ref{thecuves}) at the mass $M_B^o$ and at fixed $\gamma'$. To underline the effect due to $d$, the same procedure may be adopted for each $M_B$ along the curves: $m=10,1,0.5$ of Fig.3 by freezing the $\gamma'$ distribution to that of the reference curve $m=m^o=4$ and changing only $d$ from $0$ to the value for which at $M_B^o$ (i.e., $\gamma'_o=2.103$) $r$ becomes equal $0$ ($d=0.897$).
The results are shown in Fig. 4.
\begin{table}
\caption{Values of the intrinsic dispersion (\ref{sspr})wheighted by the two terms in $d$ and $\gamma'_o$ rigorously calculated as function of $d$ starting from the case : $m^o=4,\gamma'=2.103$ ( $r=0$ for $d=0.897$  instead of $d=1$ when $\gamma'=2$). The {\it up} and {\it down} values are obtained for variation of $m$ from $m^o=4$ up to 10 and down to 1.}
\centering
\begin{tabular}{|c|c|c|c|c|c|}
\hline
\hline
$d$ &$-\frac{1}{3-d}$&$\frac{1}{\gamma'_o}$& $(\Delta log r_e)^{up}$ &$(\Delta log r_e)^{down}$&$(\Delta log r_e)_{mean}$\\
\hline
0.000&-0.334&0.476&0.06&0.13&0.10\\
0.500&-0.400&0.476&0.03&0.07&0.05\\
0.897&-0.476&0.476&0.00&0.00&0.00\\
\hline
\end{tabular}
\label{tab:const5}
\end{table}
However, by relaxing the previous constraint, the intrinsic scatter also by varying $\gamma'$, reaches its maximum for a DM homogeneous distribution (d=0) and its minimum for a {\it cuspy} halo with $d=1$ (Fig. 5). That connection with DM {\it cusp/core} problem (see, Bindoni, 2008) is not new. In the CVMT indeed the {\it tilt} of galaxy Fundamental Plane is obtained only with a {\it cored} DM distribution ($d\simeq 0.5$) together with the explanation of many scaling relationships (see, SB9). On the contrary the observed {\it tilt} would disappear for {\it cuspy} halos with $d=1$. 


\section{Kormendy relation}

Without ambition to treat here one of the main projections of FP, we underline what is the $(I_e-r_e)$ relation we may immediately obtain from the mass-size relationship found.
Indeed the CVMT leads to:
\begin{equation}
M_B/L\sim M_B^{\alpha};\ \alpha= \frac{1-d}{3-d}
\end{equation}
From that, to get a Kormendy relation in the simplest way when: $m=4$, $d=0.5$, we use the corresponding linearized mass-size relation of the set (\ref{uslirp}):
\begin{equation}
\label{Korm}
r_e\sim (L^{\frac{1}{0.8}})^{0.755}\sim (I_e r_e^2)^{0.944}
\end{equation} 
which gives: 
\begin{equation}
\label{Kormn}
I_e\sim r_e^{-0.940}
\end{equation}
not too far ($\simeq 24\%$) from what has been deduced from observations (see, D'Onofrio et al. 2006): $$I_e\sim r_e^{-1.24}$$

Many improvements may be done in this approach taking into account the two mass regimes and/or the curves at varying $m$ instead of the linear approximation.

To be noted that, even if the slope of mass-size relationship, i.e., the value $p=0.755$, is the same for any $d$,  nevertheless the result is depending on $d$ via the {\it tilt } value of $\alpha$. So, in the case of a {\it cuspy} halo, $d=1$, i.e., corresponding to a not-tilted FP, the Kormendy's exponent of (\ref{Kormn}) turns to be $-0.675$ instead of $-0.940$ obtained with a {\it cored} halo of $d=0.5$ which is  far of about 45\% from the observed value.

\section{The m-dispersion on other scaling laws}

The scaling laws for the three main quantities of FP: $r_e,I_e, \sigma_o$ (\ref{re}, \ref{ie}, \ref{sig}) allow us to obtain not only the link between: $M_B/L-\sigma_o$ and $L-\sigma_o$ (Faber-Jackson, FJ) but also to investigate the role of m-dispersion on these main scaling relationships and exspecially on the tilt of FP.
Indeed, according to the CVMT the FJ-relation turns out to be:
\begin{equation} 
\label{FJII} 
L\sim m^{2\frac{(3-d)-\gamma'}{(3-d)^2(\gamma'-1)}}
~\sigma_o^{\frac{4\gamma'}{(\gamma'-1)(3-d)}} 
\end{equation} 

It is to be noted that both the exponents either of  $M_B$ or of $m$ are directly related to the cosmological perturbation spectrum, via $\gamma'$, and then they are depending on the amount of DM, but they are also function of how DM distributes itself inside the halo, via the exponent $d$. So we have again to handle with a scaling law which, theoretically speaking, is not a straight line in the mass range of an observed sample of galaxies, but a curve having an intrinsic dispersion due to $m$.
If we assume for a typical $DM$ halo of $M_D\simeq 10^{11}M_{\odot}$, that the typical mean value of $\gamma' \simeq 2$ ($\Lambda$CDM scenario: $n_e\simeq-2$), then (\ref{FJII}) becomes:
\begin{equation}
L\sim m^{j(d)}\sigma_o^{J(d)}~~;
\end{equation}
\begin{eqnarray}
\nonumber
d=0\div 1 \Longrightarrow 
\left\{ 
\begin{array}{l}
j=0.44\div0 \\
\\
J=2.7\div4 
\end{array} 
\right.
\end{eqnarray}
If $d=0.5$, we obtain: $$L\sim m^{0.16}\sigma_o^{3.2}.$$
It should be underlined that if $d=1$ it turns out: $$L\sim \sigma_o^4$$ without dependence on $m$. 
The range resulting for the $\sigma_o$ exponent $J$ appears to be in good agreement with the value found by Faber (1987), $2.61\pm 0.08$ (in $B$ band) and with the significantly steeper slope found by Pahre et al. (1998), $4.14\pm 0.22$ (in K band).  
So the theory predicts an intrinsic different scatter in every FJ relationship due to the role of the factor $m^j$ which disappears as soon as the FP tilt disappears ($d=1\Rightarrow \alpha=\frac{1-d}{3-d}=0$).

Moreover the scaling laws (\ref{re}, \ref{ie}, \ref{sig}) yield: 
\begin{equation}
\label{grad} 
M_B/L\sim M_B^{\alpha}\sim m^{\alpha\frac{(3-d)-\gamma'}{(3-d) (\gamma'-1)}}
\sigma_o^{\frac{2\alpha \gamma'}{\gamma'-1}} 
\end{equation} 
which in turn gives for the gradient:
\begin{equation} 
\label{gradi} 
\frac{\partial log(M_B/L)}{\partial log \sigma_o}=  \frac{2\alpha
\gamma'(M_D)}{\gamma'(M_D)-1} 
\end{equation} 

It should be noted that, due to the proportionality of $M_B$ with $M_{dyn}$ (including the fraction of DM inside $a_t$, see SB9), the same result holds for the $log (M_{dyn}/L)$ gradient.  
If again $d=0.5$, then $\alpha=0.2$, and if  $\gamma'=2$, the gradient of eq.(\ref{gradi}) becomes $0.80$, without
regarding the factor $m^{0.04}$. It is to be underlined that again for $d=1$ together with the tilt the intrinsic dispersion in the relation (\ref{gradi}) disappears. A comparison test with the same gradient, as derived from observations has been performed. Cappellari et al. (2006) found $\simeq 0.8$ for an observed sample of either fast rotators or non-rotating ETGs and S0 galaxies.
The fitted  value by J\o rgensen (1999) turns out to be $0.76\pm 0.08$. Both are in good agreement with the theoretical result.  

On the contrary the ratio $L/M_B$ which produces the tilt is totally independent of the cosmic perturbation spectrum and of the mass ratio $m$. It turns out to be dependent only on the dark matter density profile. Indeed, by definition it is: 
$$L/M_B\sim I_er_e^2/M_B$$
and by remembering the relationship between $r_e\sim a_t\sim a_D$ given by eq.(\ref{at}), the following functional dependence on the bright mass holds:
\begin{equation} 
\label{iecosm} 
L/M_B\sim M_B^{(\frac{2}{3-d}-\frac{2}{\gamma'})}
M_B^{\frac{2}{\gamma'}}M_B^{-1} \sim M_B^{-1+\frac{2}{3-d}} 
\end{equation}
where the factor $m^{2\frac{\gamma'-(3-d)}{\gamma'(3-d)}}$ introduced by $I_e$ is perfectly compensated by the factor
$m^{2\frac{(3-d)-\gamma'}{\gamma'(3-d)}}$ which enters $r_e^2$ (see (\ref{re}) and (\ref{ie})) whichever is the value of $d$.
That is relevant into the explanation of the {\it tightness} of FP.

Eq.(\ref{iecosm}) tells us how the ratio $L/M_B$ loses its direct connection with the cosmology given by $\gamma'$. This confirmes the observation just made by Djorgovski (1992) that inside the FP as a whole there is a degeneracy in respect to the cosmological process of galaxy formation which breaks into the projections of FP. Indeed  the $\gamma'$ springs out by considering what FP  produces into the conjugate planes: i.e., the Faber-Jackson and the Kormendy relation.

The above considerations are restricted to the effect of interaction energy between DM and star subsystem on the FP.   Other physical ingredients induce surely a spread, such as the parameters of stellar populations, i.e. age and
metallicity, rotational kinetic energy, radial orbital anisotropy and projection effects due to lack of spherical symmetry (e.g., Saglia et al., 1993). A detailed investigation on this point is outside the aim of the current paper. 

\section{Discussion}

Before conclusions we have to take briefly into account: i) the relationship of CVMT, in which two-component virial equilibrium holds conserving {\it homology}, while the {\it weak homology} appears in the FP literature (see, e.g., Bertin et al., 2002); and ii) the consistency of CMTV interpretations in front of different new sets of data.

i) It is important to distinguish between the dynamical effect the TCVM is able 
to offer in order to explain the tilt of FP without breaking the homology 
and the description of how  the weak homology holds by adopting Sersic 
luminosity profiles.
We mean: the dynamical theory tells us why the $M/L \sim M^{\alpha}$ holds but 
it is not able to enter the details of how the light is distributed on the 
galaxy surface. On the other side, Sersic luminosity profiles are able to describe how is the best 
distribution of light (e.g., Kormendy et al., 2009) without explaining us why the total light $L$ has to 
change by increasing $M$, i.e., $L\sim M^{1-\alpha}$.
They appear as two complementary aspects of the main problem which is the 
change of light with mass. The description of Sersic does not enter the 
cause of the phenomenon, the explanation of the phenomenon does not enter 
the detailed description of its consequence.

ii) We refer to the huge amount of database SDSS, published on the paper of Bern11. There are at least two very interesting outputs which spring out from the new relationships here derived by fitting the data. The first one is the correlation obtained for size and velocity dispersion $\sigma$ (see, Bernardi et al., 2003a). From Fig. 2 of Bern11 the correlation turns to be: $ r_e\sim \sigma^{1.5}$ in the mass range $log M_*=10.5\div 11.3$. Taking into account the Eqs. ((\ref{re}) and (\ref{sig})) of the scaling laws, the prevision of the theory for the $\sigma_o$ exponent, in the same mass range, in the case of $m_o=4$, disregarding the curvature of the mass-size relation, is: $2R/(1-R)\simeq 1.8$.
The second one is connected with the slope of the mass-size correlation once extended to the low mass regime.
In the present paper we may evaluate the local slope down to $(M_B/M_{\odot})^{9.60}$. Disregarding the curvature it turns to be: $p=1/\gamma'= 0.437$ in the case $m_o=4$ in agreement with Bern11 (Fig.1) where it has to be less than about $0.55$.     

\section{Conclusions}

By Clausius' virial maximum theory we have derived a $M_*-r_e$ theoretical relationship for ETGs in a $\Lambda$CDM cosmological scenario which has been compared with that found by fitting the binned averaged data of PS96 (Tal09, Fig.4). It is to be underlined that our theoretical slope changes with $M_*$ and increases with it. That is due to the increasing of the local slope in the cosmological mass variance at increasing DM halo mass. The connection between cosmological scenario and luminous {\it mass-size} relationship is one of the best result of the CVMT theory. Inside it many {\it mass-size} relationships are possible depending on the given Dark/Bright mass ratio $m$. The corresponding curves may be linearized in some fruitful numerical relations with a maximum scatter in respect to the right curves less than $1.5$ times the rms observational errors ($m=1\div 4$). 
Assuming the existence of two mass regimes, upper and lower $10^{11}\ M_{\odot}$ according to Tal09, the agreement at lower regimes turns to be unsatisfactory, probably due to the crossing of different {\it mass-size} relationships  at different $m$ in the numerical fit of Tal09. At higher regimes the agreement turns to be fairly well. The difference between lower and higher regimes, underlined by Tal09, seems to be not so relevant from the theoretical point of view, due to the gradually increasing of the slope passing from one regime to the other without solution of continuity. Moreover, mass ratios $m$ around $1-4$ are more suitable to the galaxy sample used by Tal09, while values of about $10$ appear to be ruled out, according to a $\Lambda$CDM scenario. 
 
The existence of individual galaxies with different values of $m$ actually produces a real scatter in the data we can see in the non-binned points (Fig.4 of Tal09). This precious information is then locked inside the binning procedure. 
It is relevant that the amplitude of this dispersion is to connect with the dark matter inner distribution and more precisely that it increases or decreases going towards one {\it cored} or {\it cuspy}, respectively. 
 
The last but not the least consideration has to be drawn about the capability of CVMT: the spread in the single galaxy data appears to be limited better from the lower part of the ($M_*-r_e$) plane in respect to the higher one. The CVMT tells us that there has to exist a minimum value of $m$ in the sample due to the presence of a sharper border on the side of lower $r_e$. But what is its physical meaning? We know (Bindoni, 2008) that as soon as $m$ decreases the ratio $x=a_B/a_D$ increases towards $1$. A limit $m_l$ exists and corresponds to $a_B=a_D$. After this limit the CV maximun can not exist because the equipartion of Clausius' virial energy is no longer possible. That limit value is $m_l\leq 1$ and has to corresponds to the ZOE in the $k-space$ rapresentation of FP. A reflex of ZOE is then also present inside the {\it mass-size} relationship with a "`Zone of Avoidance"' delimited by the $m_l-line$ (e.g., D'Onofrio et al., 2006).

Many problems are still open in our approach. For example, the assumed homology requests to choose one common value for the King's concentration $C_B$ and a unique CV maximum in the whole galaxy family to assure the same theoretical FP which implicitly underlies. But it is important to explore what happens by changing them. 
 
As general comment we can underline that the message which tells us this $mass-size$ relationship is again that the true challenge is not to interpret in some way it but to understand why there are so many scaling relations of this kind for galaxies. In our opinion that is impossible to do whithout the cosmological framework because their existence is stricktly connected with how galaxies form. The Clausius' virial maximum theory included this link by nature.
Based on classical physical grounds here revised the CVMT looks like very promising in order to understand the general features related to all astrophysical virialized structures and collected by Burstein et al. (1997) as {\it cosmic metaplane}.   


 \section*{Acknowledgements}
      Thanks are due to C. Tortora for fruitful discussions and
      additional data on his (and coauthors) quoted paper Tal09.
      Deep gratitude is expressed to an anonymous referee for enlightening
      comments.


\end{document}